\documentclass[preprint,12pt]{elsarticle}

\usepackage{epsfig}
\usepackage{epstopdf}

\usepackage{amssymb}
\usepackage{amsthm}

\newcounter{bla}

\journal{Computer Physics Communications}

\begin{document}

\begin{frontmatter}



\title{PHOTOPiC: Calculate photo-ionization functions and model coefficients for gas discharge simulations}


\author[a,b]{Yifei ZHU\corref{author}}
\author[a,b]{Yun WU\corref{author}}
\author[c]{Jianzhong LI}

\cortext[author] {Corresponding author.\\\textit{E-mail address:} yifei.zhu.plasma@gmail.com,wuyun1223@126.com}
\address[a]{Science and Technology of Plasma Dynamics Laboratory, Airforce Engineering University, Xi'an 710038, People's Republic of China}
\address[b]{Institute of Aero-engine, School of Mechanical Engineering, Xi’an Jiaotong University, Xi’an 710049, People's Republic of China}
\address[c]{Gongfang Plasma Tech. Co. Ltd. Hefei, China}

\begin{abstract}

A program to compute photo-ionization functions and fitting parameters for an efficient photo--ionization model is presented. The code integrates the product of spectrum emission intensity, the photo--ionization yield and the absorption coefficient to calculate the photo--ionization function of each gas and the total photo--ionization function of the mixture. The coefficients of Helmholtz photo--ionization model is obtained by fitting the total photo--ionization function. A database consisting $\rm N_2$, $\rm O_2$, $\rm CO_2$ and $\rm H_2O$ molecules are included and can be modified by the users. The program provides more accurate photo--ionization functions and source terms for plasma fluid models. 

\end{abstract}


\end{frontmatter}



{\bf PROGRAM SUMMARY}

\begin{small}
\noindent
{\em Program Title: \rm PHOTOPiC}                                          \\
{\em Licensing provisions: \rm GPLv3 }                                   \\
{\em Programming language: \rm Python}                                   \\
{\em Nature of problem: \rm Photo--ionization plays a critical role in creating seed electrons in front of the ionization streamer head. The photo--ionization functions are required to calculate source terms in fluid models for pure gas or gas mixtures. The Efficient three--terms Helmholtz photo--ionization model has to be extended for non air gases but the modeling parameters are unknown.}\\
{\em Solution method: \rm the code integrates the spectrum emission intensity, the photo--ionization yield and the absorption coefficient to calculate the photo--ionization function of each gas\cite{pancheshnyi2014photoionization} and the total photo--ionization function of the mixture. The coefficients of Helmholtz model is obtained by fitting the total photo--ionization function. }\\
{\em Restrictions: \rm The included database is not complete. The users have to provide by themselves the quenching pressures and ionization efficiency. The methods have been introduced in the text.}\\
{\em Running time: \rm Few seconds for pure gas and few minutes for a gas mixture of two or three species.}\\

\end{small}

\section{Introduction}

During the gas discharge, effective excitation of radiative states occurs, and the radiation propagates outward, producing photoelectrons at various distances from the discharge zone. As a result, electrons can be produced at a certain distance from the discharge propagating head, known as photo--ionization.
Photo--ionization plays a critical role in the spatial advancement of discharge streamers by creating seed electrons in front of the ionization head~\cite{morrow1997streamer,Pancheshnyi2001,Naidis2006,liu2006effects,nudnova2008streamer,nijdam2010probing,wormeester2010probing, Naidis2018}.

The photo--ionization rate of a given gas or gas mixture can be characterized by the pressure and geometry independent photo--ionization function. Notwithstanding the increasing demand for photo--ionization rates, there are still only very few publications on this topic. A recent study can be found in Ref~\cite{pancheshnyi2014photoionization}, where the photo--ionization functions of $\rm N_2$, $\rm O_2$, $\rm CO_2$ and air are given and discussed. The influence of water addition to the air has been discussed in \cite{Naidis2006}.

Numerical simulation of gas discharges requires accurate and efficient evaluation of the effects of photo--ionization, which remains a challenging task. To account for photo--ionizations, the continuity equations of electrons and ions in the plasma fluid model have to be added with a source term~\cite{Bourdon2007}. With the knowledge of photo--ionization functions, the photo--ionization source term can be calculated by the classical integral model~\cite{Zhelezniak1982,Pancheshnyi2001,xiong2012dynamics}, the two or three terms Helmholtz models~\cite{Luque2007,Bourdon2007}, the three-group Eddington and SP$_3$ approximation model~\cite{segur2006use,Bourdon2007}, the Monte-Carlo model~\cite{Teunissen_2017} et al.

The aforementioned models for photo--ionization source terms are considered mainly in simulations of gas discharges in $\rm N_2$ and $\rm O_2$ mixtures. For example, the three terms Helmholtz model is one of the widely used approaches~\cite{liu2007application,likhanskii2009role,papageorgiou2011three,breden2012self,zhuang2014weno,zhu2017nanosecond,zhu2018fgh,Sharma2018,zhihang2020two}: it is easy to implement and requires only a Poisson solver, but existing coefficients presented in \cite{Bourdon2007} are applicable only for $\rm N_2/\rm O_2$ mixtures and the concentration of $\rm O_2$ should not be zero, otherwise will loss accuracy due to changed photo--ionization functions~\cite{wormeester2010probing}. 

In this work we present the PHOTOPiC (PHOTO--ionization Parameter Calculator) program to calculate the photo--ionization function of pure gas or gas mixtures (with known photo--ionization and photo--absorption cross--sections), and an improved three--terms Helmholtz model to account for photo--ionization in non $\rm N_2/\rm O_2$ gases.

\section{Theoretical background}

\subsection{Photo--ionization yield and absorption coefficient}	
The probability of photo-ionization is related to the photo-ionization cross-section, which depends on the energy of the photon and the target being considered. The probability of absorbing radiation is related to the absorption cross-section. EUV absorption and photoionization cross sections for the gases under consideration can be found in different publications~\cite{hudson1971critical,fennelly1992photoionization}. The combination of these parameters gives the information of how the photo emission will be absorbed by each species in the system.

At each wavelength $\lambda$ of individual gas specie $i$, the photo-ionization yield $\xi_{\lambda,i}(\lambda)$ (the probability of ionization of molecules by photon absorption) and absorption coefficient $\mu_{\lambda,i}(\lambda)$ are defined as:

\begin{equation}
\xi_{\lambda,i}(\lambda)=\frac{\sigma_{ion}(\lambda)}{\sigma_{abs}(\lambda)}
\end{equation}	

\begin{equation}
\mu_{\lambda,i}(\lambda)/p=\sigma_{abs}(\lambda)\frac{1}{k_BT}
\end{equation}	
\noindent
where $k_B$ is the Boltzmann constant and $T$ is the gas temperature. $\sigma_{abs}(\lambda)$ and $\sigma_{ion}(\lambda)$ are photo absorption and photo--ionization cross sections at specific wavelength. Above values are essential to calculate the multi--component photo--ionization functions.

\subsection{The photo-ionization functions}
In the widely used model derived by Zheleznyak et al~\cite{Zhelezniak1982,liu2004effects,Naidis2018} for photoionization in air, the photo--ionization rate at point of observation $\vec r_1$ due to source points emitting photo--ionizing UV photons at $\vec r_2$ is:

\begin{equation}
S_{ph}(\vec r_1)=\int_V \frac{I(\vec r_2)\Phi_0(r)}{4\pi r^2} dV
\end{equation}	
\noindent
where $r=|\vec{r_1}-\vec{r_2}|$. For simplicity $I(\vec{r})$ is assumed to be the ionization production rate. The photo--ionization function $\Phi_0(r)$ reduced by pressure can be expressed by following formula~\cite{pancheshnyi2014photoionization}:

\begin{equation}\label{eqs_singlefunction}
\frac{\Phi_0(r)}{p}=\Pi\frac{1}{4\pi}\cdot\frac{\omega}{\alpha_{eff}}\cdot\frac{\int_{\lambda_{min}}^{\lambda_{max}}\xi_{\lambda}(\mu_{\lambda}/p)exp(-(\mu_{\lambda}/p)pr)I_0d\lambda}{\int_{\lambda_{min}}^{\lambda_{max}}I_0d\lambda}
\end{equation}
\noindent
where $\Pi$ is the correction factor for high pressure conditions, $\omega/\alpha_{eff}$ is the scaling factor. The integral range $(\lambda_{min},\lambda_{max})$ is decided by the threshold ionization energy and the dominating radiation transitions at higher energy levels. $I_0d\lambda$ is the emission spectrum of the gas intensity at given wavelength and electron energy.

Equation (\ref{eqs_singlefunction}) is defined for only one specie. In case of a multi--component system, the concentration of the absorbing specie $j$, $\eta_j$ ,has to be taken into account to calculate the absorption of radiation by all species in the system. In equation~(\ref{eqs_singlefunction}) the absorption term in $exp(-(\mu_{\lambda}/p)pr)$, $\mu_{\lambda}$ is redefined by:

\begin{equation}
\mu_{\lambda}^{ij}=\sum_{j}\eta_j\mu_{\lambda,j}
\end{equation}

When the contribution from emission and absorption of every specie are considered, for gas specie $i$ ionized by emission from specie $j$, the multi--component photo--ionization function reads:

\begin{equation}\label{eqs_multifunction}
\frac{\Phi_0(r)_{ij}}{p}=\eta_j\frac{1}{4\pi}(\frac{\omega}{\alpha_{eff}})_{ij}(\frac{\int_{\lambda_{min}}^{\lambda_{max}}\xi_{\lambda,i}(\mu_{\lambda,i}/p)exp(-(\mu_{\lambda}^{ij}/p)pr)I_{\lambda,j}^0d\lambda}{\int_{\lambda_{min}}^{\lambda_{max}}I_{\lambda,j}^0d\lambda})\Pi_{ij}
\end{equation}

In a gas system of $n$ species, there will be $n\times n$ photo-ionization functions. These functions can then be used to analyze the constitute of photo-ionization source. This enables the possibility to find the main contributing photo--ionization processes to the gas discharge propagation. 

\subsection{Parameters for the extended Helmholtz photo--ionization model}
The three terms Helmholtz photo-ionization model has been widely used in many groups as it is easy to be implemented and requires only a Poisson solver. The three terms Helmholtz model writes:

\begin{equation}\label{eqs.photoionization}
\nabla^2S_{ph}^j(\vec{r}) -(\lambda_jp)^2S_{ph}^j(\vec{r})=-A_jp^{2}I(\vec{r})~~(j=1,2,3)
\end{equation}

The sum of $S_{ph}^j$ is the source term of photo-ionization, $\lambda_j$ and $A_j$ are parameters to fit. 
It has to noted that, in previous publications $p$ denotes the partial pressure of oxygen~\cite{Bourdon2007} and $I$ is the product of ionization source, scaling factor and pressure correction factor. The classical three-terms Helmholtz model are applicable only for N$_2$:O$_2$ mixtures (and O$_2$ molecules must be the ionizing target)\cite{wormeester2010probing}. The use of this model for other gas mixtures haven't been found yet. 

In this work, we consider $p$ as the gas pressure, and $I(\vec{r})$ is the ionization source term, allowing a more general condition. The parameters $\lambda_j$ and $A_j$ follow:

\begin{equation}
\frac{\Phi_0(r)}{p}=(pr)\sum_{j}A_je^{-\lambda_jpr}
\end{equation}

Taking $pr$ as the variables, and knowing the value of photo--ionization function $\Phi_0(pr)/p$, the least square method is used to find the fitting parameters. Typically in low temperature plasma streamer stage the gas mixture is not changing dramatically, thus we can use the fitted parameters as constants in the simulation.

\section{Program organization}

The program PHOTOPiC computes the photo--ionization functions by integrating the products of photo--ionization yields, absorption coefficients and emission spectrum intensity. It requires the input of three formatted files containing the photo--ionization and absorption cross sections, the emission spectrum data and other physical parameters (wavelength range, the scaling and quenching factors).

When running, the program collects the information of gases (species, concentrations, temperature and pressure), reads input files (cross sections, emission spectrum and other physical parameters) and then calculates the photo--ionization functions of individual gases and sums to achieve the total photo--ionization function. The parameters of three terms Helmholtz photo--ionization model are then calculated based on fitting of the total photo--ionization functions.

\subsection{Distribution and installation of the program}
The program is distributed as a single zip file containing an executable, three input files (Xsecs.dat, Spectrum.dat, Phys.dat) and the source code. At least 80~Mb hardware space are required. The executable can be ran directly on a 64~bit windows operation system.

It is also possible to run the program from the source code. A Python~3.7 environment with corresponding packages listed in the requirements.txt have to be installed by pip command if the users wants to run the program from client.py in the unzipped folder. 

\subsection{Graphical user interface}
In this section the usage of the program is demonstrated, taking calculation of the photo--ionization functions of air as an example. The program starts with the gas specification page. One has to select the gases to be studied, and define the concentration ($\%$). The sum of the percentage of each concentration should be 100$\%$.

\begin{figure}[t!h!]
	\epsfxsize=\columnwidth
	\begin{center}
		\epsfbox{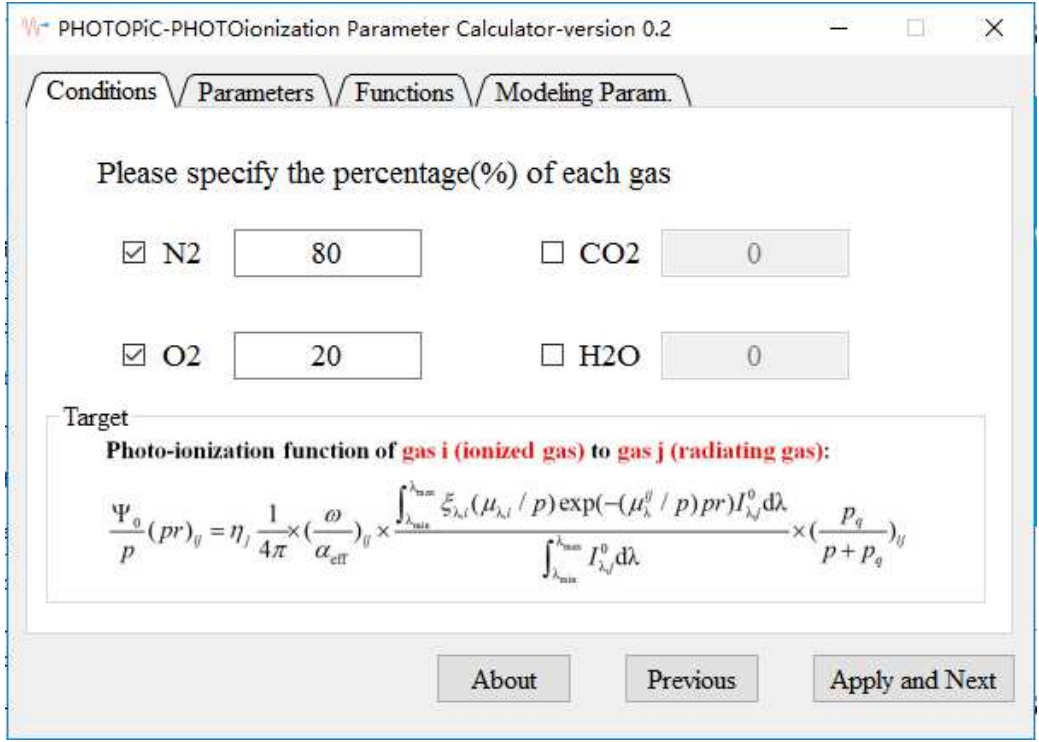}	
		\caption{Specify the components and concentration of a gas mixture.}
		\label{page1}
	\end{center}
\end{figure}

\begin{figure}[t!h!]
	\epsfxsize=\columnwidth
	\begin{center}
		\epsfbox{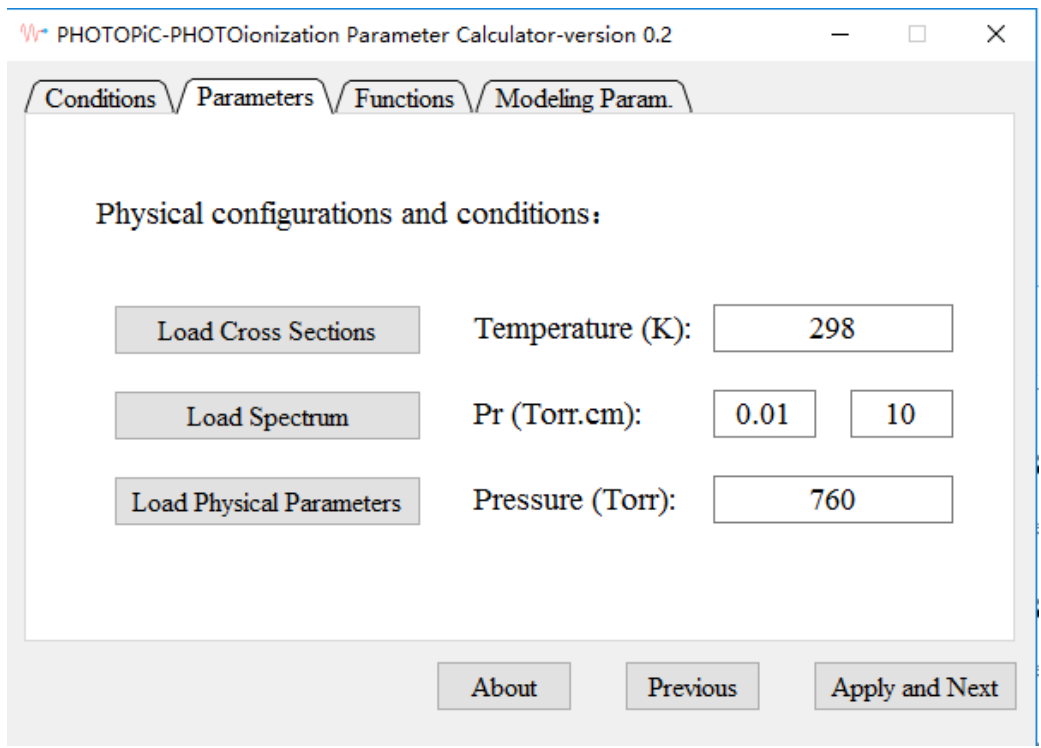}	
		\caption{Specify the condition and physical parameters required to calculate photo--ionization function.}
		\label{page2}
	\end{center}
\end{figure}

Figure~\ref{page1} shows the default settings. One can click the ``apply and next'' button after filling all the ticked text boxes. The users can tick as many species as they want, but the increase number of specie means longer calculation time. If the concentration is 0$\%$, just tick out to reduce the computational cost. After specifying the mixtures, users have to click ``Apply and Next'' to proceed to next stage and designate physical parameters, see Figure~\ref{page2} . 

The condition parameters required for photo-ionization function calculation are temperature (K), Pr (the product of pressure and radius, Torr$\cdot$cm) and pressure (Torr), these can be specified directly in the graphical user interface.

The physical parameters as input are photo-ionization cross sections, photo-absorption cross sections, emission spectrum, integrating range $\lambda_{min}$, $\lambda_{max}$, the scale factor $\omega/\alpha_{eff}$ and the quenching pressure $p_q$ for the pressure correction factor $\Pi$. We have organized three database files containing those parameters for $\rm N_2$, $\rm O_2$, $\rm CO_2$ and $\rm H_2O$ gases. The program will automatically search and load the input database files: Xsecs.dat, Spectrum.dat and Phys.dat in the program folder, if the files are absent, the users have to load manually by clicking on the buttons and select corresponding files. The database files can be updated or replaced with other gases, the format of the input files will be introduced in the next section.

After specifying the condition and physical data, users can click  ``Apply and Next'' to proceed to the next stage and calculate the photo-ionization function of each gas ionized by the radiation from another. 

\begin{figure}[t!h!]
	\epsfxsize=\columnwidth
	\begin{center}
		\epsfbox{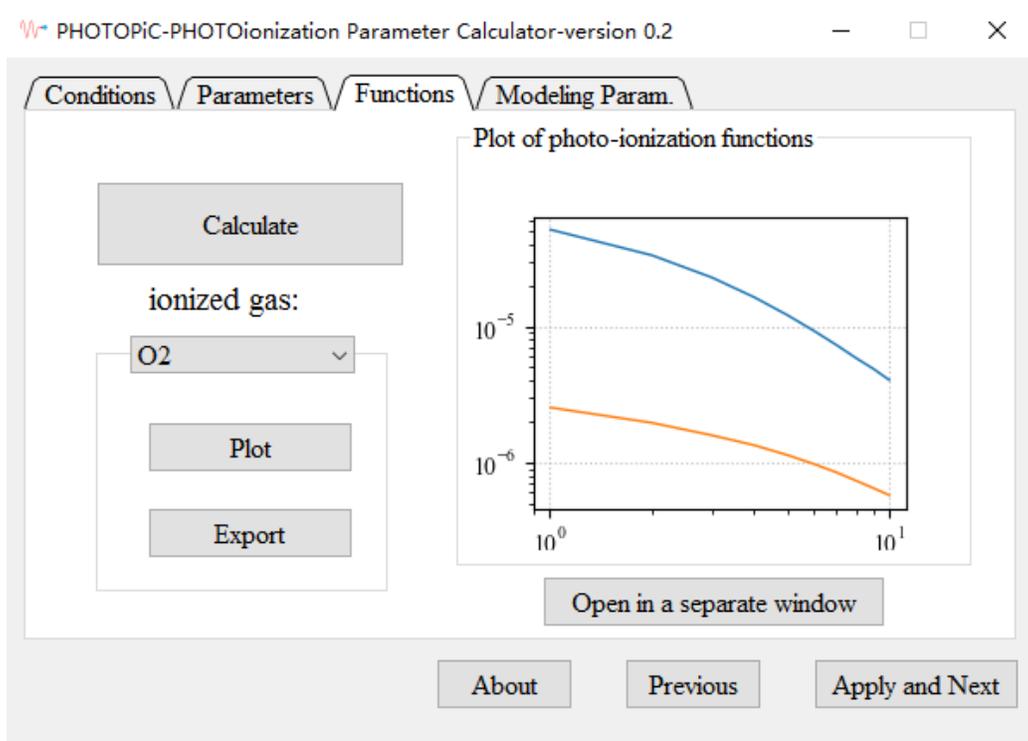}	
	\caption{Label-1 specify a gas and corresponding percentage}
\label{page3-3}
	\end{center}
\end{figure}

In this stage the users just have to click the ``Calculate'' button and wait. When PHOTOPiC is calculating, all the button will be grey, there will be a progress information on the left corner. Once the calculation is finished, the grey buttons will turn active (black) again. Then the users can plot the results in the window on the right, as shown in Figure~\ref{page3-3}. To plot, first select from the list ``ionized gas'' the target gas specie, and then click the ``Plot'' button. You will see the photo--ionization functions caused by different source gases in the window on the right. The results can be saved by clicking the ``Export'' button.

It has to be noted that sometimes the plot window is too small to show enough information (legends et al.). Users can click the ``Open in a separate window'' button, to see the results in a full--screen window or just treat the data in the output file.

\begin{figure}[t!h!]
	\epsfxsize=\columnwidth
	\begin{center}
		\epsfbox{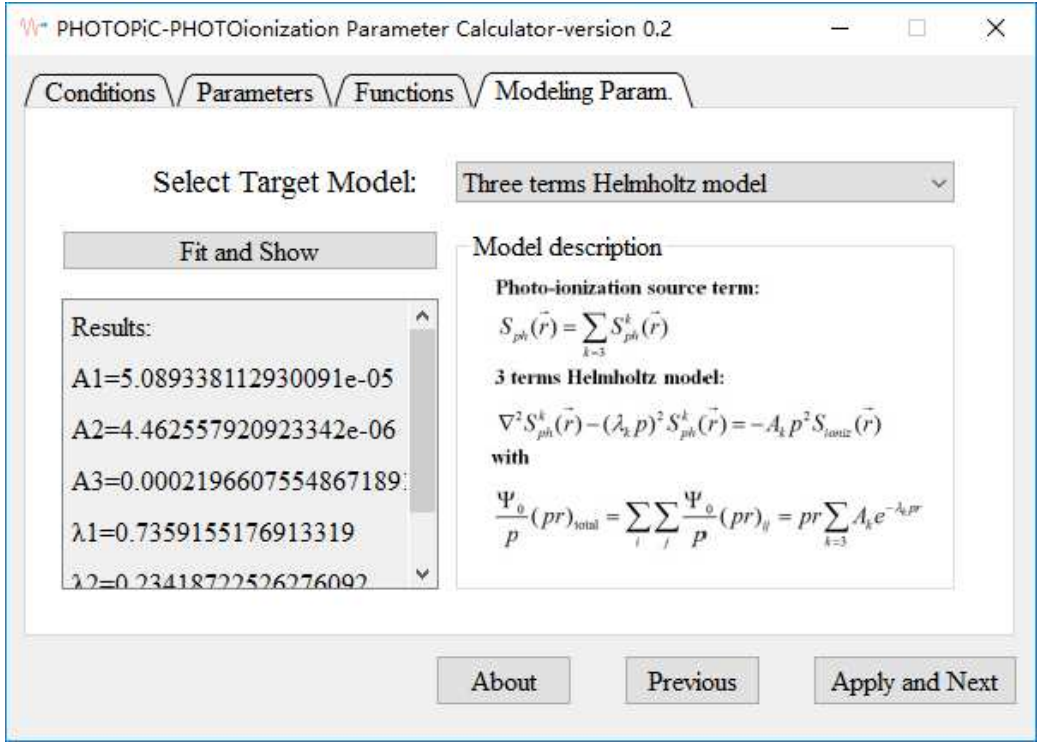}	
	\caption{Label-1 specify a gas and corresponding percentage}
\label{page4}
	\end{center}
\end{figure}

To bridge the photo--ionization and the models for the calculation of photo--ionization, one has to click the ``Apply and next'' to proceed to next stage. This stage parameters for specific models that are widely used in plasma modeling (especially streamer discharges) can be calculated. This first version of the software calculates only for the extended 3--terms Helmholtz equations model.

To obtain the parameters, just select the model from the list (see Figure~\ref{page4}), the description of the model is shown below the list. Click the ``Fit and show'' button. The fitting parameters are calculated using the least square method and printed in the text box below the button.

The fitting parameters can be used in fluid codes or existing commercial softwares (for example COMSOL multiphysics) with tiny modifications to the classical three terms Helmholtz photo--ionization equations (replace the oxygen partial pressure with the pressure).

\section{Input data specification}
In this section the format of the input files are explained. The users can adjust the database files to replace the gases to calculate any other specific cases.
\\
\\
(1) cross sections

Photo--ionization cross section and Photo--absorption cross section data are stored in Xsecs.dat file. The cross section data in the database is extracted from~\cite{fennelly1992photoionization,Cairns1965,Cairns1965}. The template of the database is:
\\
\\
$\#$Gas name\\
$\#$photo-ionization or photo-absorption\\
Energy(eV)	xsec($\times10^{-18} m^2$)\\

Take N$_2$ as an example, the cross section data reads:
\\
\\
$\#$N2\\
$\#$photo-ionization\\
2.370	0.80\\
2.847	0.96\\
2.879	0.95\\
2.952	0.99\\
3.000	0.61\\
...\\
$\#$N2\\
$\#$photo-absorption\\
2.370	0.80\\
2.847	0.96\\
2.879	0.97\\

The program assumes that the first column of the photo--ionization and photo--absorption cross sections are the same (as they are usually provided together). It is suggested that the users make an interpolation manually to meet this restriction, if they are providing the cross section data from different sources with different discrete values in the first column.
\\
\\
(2)Optical emission spectrum

Optical emission spectrum in an as large wavelength range as possible for each gas is suggested. The database of the spectrum is stored in Spectrum.dat file. The data file can be found in existing publications~\cite{Ajello1985,Ajello1989,Kanik1993,Ajello1984}. The template of the database is:
\\
\\
$\#$Gas name\\
$\#$some comments\\
Wavelength(nm)	strength(a.u.)\\

Still, take N$_2$ as an example, then spectrum data reads:
\\
\\
$\#$N2\\
$\#$precision 0.5nm, 50--200nm\\
52.5773	0.01997\\
52.8522	0.08068\\
53.1271	0.18684\\
53.2646	0.23235\\
53.5395	0.15670\\
...
\\
\\
(3) Other physical parameters

Additional physical parameters are necessary to decide the integration limit of spectrum, the scaling factor relating to the emitting processes and the quenching factors. Usually the users don't have to adjust the incorporated database. But if users would like to replace with new gases, or to update from latest publications, it is necessary to check carefully these parameters. The list of other physical parameters are:

(i) The integration wavelength range $\lambda_{min}$ and $\lambda_{max}$. The lower and upper limit is denoted as ``LAMBDA$\_$MIN'' and ``LAMBDA$\_$MAX''.  

The upper limit corresponds to the ionization threshold energy of the target gas $j$, thus there is 4 numbers in ``LAMBDA$\_$MAX'', corresponding to N$_2$ ($j=1$), O$_2$ ($j=2$), H$_2$O ($j=3$) and CO$_2$ ($j=4$).

The lower limit is decided to ensure there are at least two peak emission intensities between ``LAMBDA$\_$MIN'' and ``LAMBDA$\_$MAX'' for the emitting gas $i$. Thus there are n$\times$n (n is the number of species in the database) values to be decided, they are tabulated as a matrix, if $i$=2, $j$=3 then ``LAMBDA$\_$MIN'' value is located at the second row and third column.

(ii) The scaling factor $\omega/\alpha_{eff}$. This factor does not affect the profile of photo--ionization functions, but decides the absolute value. It corresponds to the excitation of one system of radiative transitions and is also a matrix denoting the radiating gas $i$ and target gas $j$. This value can be obtained by two steps:

First, find the cross section of corresponding emitting reactions of gas $i$, $\sigma_{ext}$. Divide $\sigma_{ext}$ by $\chi$ (the percentage of the emitting intensity to the total emission intensity, see \cite{pancheshnyi2014photoionization}):

\begin{equation}
\sigma_{ext}=\sigma_{ext}/\chi
\end{equation}

Second, use the scaled cross section $\sigma_{ext}$	with the cross sections of all species in the mixture at wanted ratio, run a Boltzmann equation solver (e.g. BOLSIG+~\cite{Hagelaar2005}) to get the rate coefficients. Then the scaling factor can be calculated according to:

\begin{equation}
\omega/\alpha_{eff} (E/N)=\frac{k_{exc}(E/N)}{\sum k_{ion}(E/N)-k_{attch}(E/N)}
\end{equation}
\noindent
where the rate of ionization $k_{ion}$ and attachment $k_{attach}$ can be calculated from the Boltzmann equation, too.

(iii) Quenching pressure $p_q$. At elevated pressures, the quenching of radiative states has to be taken into consideration. The pressure correction factor in equation~(\ref{eqs_singlefunction}) is defined by:

\begin{equation}
\Pi=\frac{p_q}{p+p_q}
\end{equation}

The value of quenching pressure $p_q$ is defined according to stationary kinetics balance of excited radiating species by taking into account radiative decay and quenching. The value can be decided from following formulation:

\begin{equation}
p_q=\frac{k_BT}{\tau_0k_q}
\end{equation}	
\noindent where $\tau_0$ is the radiative decay time and $k_q$ is the quenching rate of corresponding emitting species. The uncertainty of $k_q$ leads to some deviations. Typical $p_q$ for pure nitrogen, oxygen and nitrogen in air is 9.8, 30 and 36 pa, respectively~\cite{pancheshnyi2014photoionization}. This value is tabulated as a matrix, the quenching factor of excited gas $i$ by gas $j$ is stored in row $i$ and column $j$ of the matrix. 
\\
\\
(4) Special cases

The input data has to be adjusted in case of following cases: 

(i) To calculate for low pressures. In this case just set a huge quenching pressure, thus $\Pi\approx1$. 

(ii) To compare with some classical formulations. That is to say, if users want to get the photo--ionization function without scaling factors and quenching factors (like in Ref~\cite{Bourdon2007} et al.), just set the scaling factor as 1. 

(iii) To calculate a specific gas that is not included in the database. To achieve this, one can select a gas (for example N$_2$) and replace the cross--section, the spectrum and corresponding physical data, ``pretending'' that $\rm N_2$ is the desired gas specie.	

\section{Examples and benchmarks}
In this section the calculated photo--ionization functions of $\rm N_2$, $\rm O_2$ and $\rm CO_2$ molecules are presented and compared with existing experimental data or estimations. The photo--ionization source term to be used in a fluid model is also calculated and compared with a benchmark case for $\rm N_2/\rm O_2$ mixture. Finally the extended three terms Helmholtz photo--ionization model is tested using the parameters provided by PHOTOPiC with a two dimensional streamer benchmark simulation.

\subsection{Photo--ionization functions of $\rm N_2$, $\rm O_2$, $\rm CO_2$ and $\rm H_2O$ molecules}
The calculated photo--ionization functions by PHOTOPiC are compared with experimental values for pure N$_2$, O$_2$ and air. The experimental values have also been summarized in paper~\cite{pancheshnyi2014photoionization}.

\begin{figure}[t!h!]
	\epsfxsize=\columnwidth
	\begin{center}
		\epsfbox{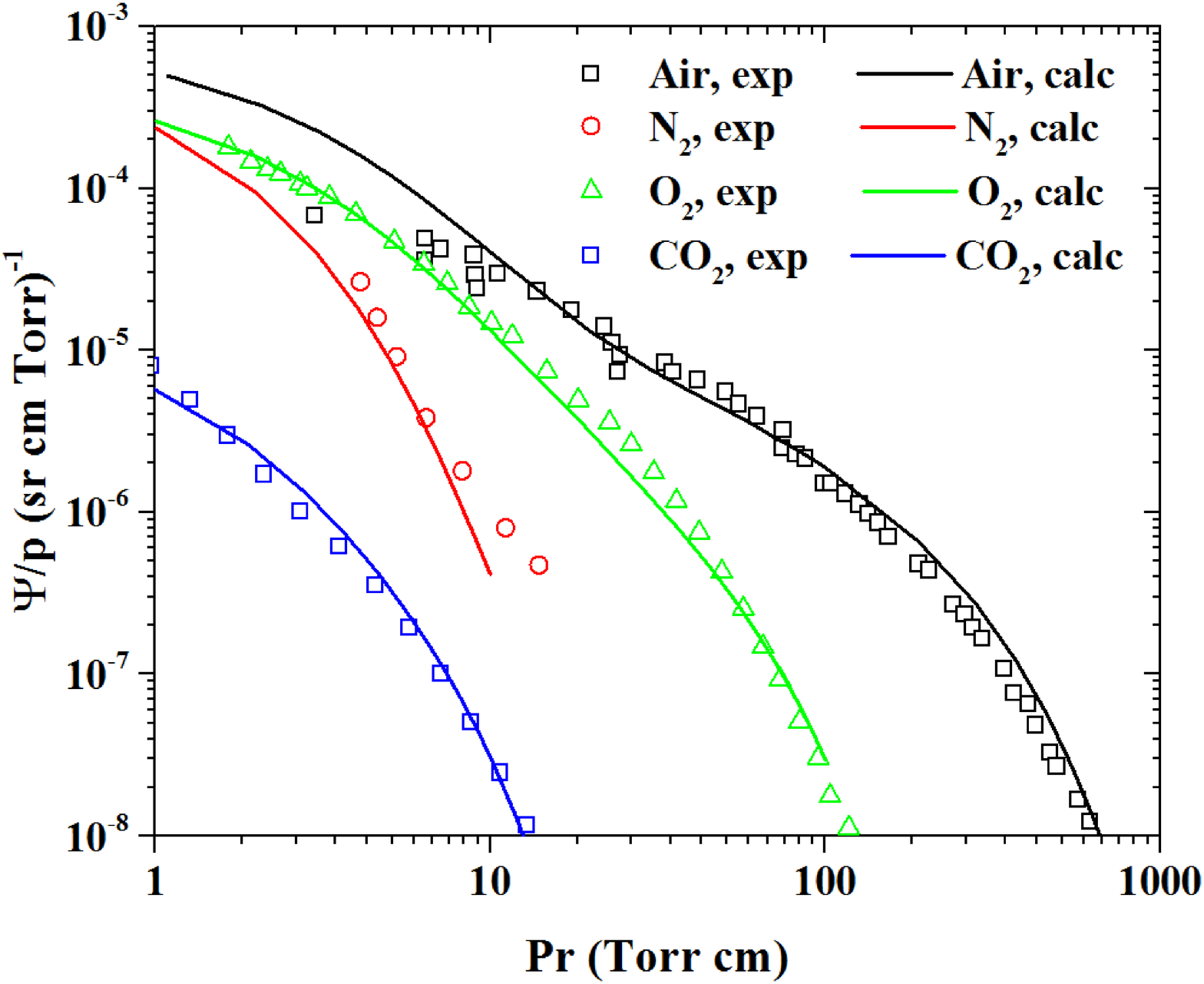}	
	\caption{Photo--ionization functions calculated by PHOTOPiC and from experimental data. Air data correspond to the work of Penney~\cite{penney1970photoionization}, O$_2$ data correspond to the work of Przybylski~\cite{przybylski1962untersuchung}, and N$_2$ data is from the work of Teich~\cite{teich1967emission}, CO$_2$ data is retrieved from Ref~\cite{pancheshnyi2014photoionization}.}
\label{validation1}
	\end{center}
\end{figure}

Using the default database, PHOTOPiC well reproduced the photo--ionization functions reported in the references~\cite{penney1970photoionization,przybylski1962untersuchung,teich1967emission,pancheshnyi2014photoionization}, as shown in Figure~\ref{validation1}. The photo--ionization function drops exponentially with $Pr$, the absolute value and decreasing rate varies. Photo--ionization function of air is the highest in absolute value and in $Pr$ range, indicating that photo--ionization in a gas mixture can be significantly enhanced, which has also been validated in \cite{nijdam2010probing,wormeester2010probing}. 

Notes that in Ref~\cite{pancheshnyi2014photoionization} the quenching factors are not taken account, thus the results obtained by PHOTOPiC has to be multiplied by $(p+p_q)/p_q$ for the comparison. 


\subsection{Photo--ionization source terms of air}

The fitting parameters for the extended Helmholtz Photo--ionization model calculated by PHOTOPiC are checked by comparing with results presented in paper~\cite{Bourdon2007}. The gas mixture for validation is air. We compared the photoionization production rate $S_{ph}$ in a two-dimensional axisymmetric domain assuming a Gaussian-distribution ionization source:

\begin{equation}
I(r,z)=I_0exp(-(z-z_0)^2/\sigma^2-r^2/\sigma^2)
\end{equation}
\noindent
where $z_0$ is the axial position of the source term, $\sigma$ controls the spatial width of the ionization source, $I_0=1.53\times10^{25} cm^{-3}s^{-1}$ is the ionization source. The calculations are performed in three different dimensions $0.02\times0.02 cm^2$, $0.2\times0.2 cm^2$ and $2\times2 cm^2$, the value of $\sigma$ in corresponding dimensions are 0.001, 0.01 and 0.1 cm, $z_0$ equals 0.01, 0.1 and 1~cm, respectively.

\begin{figure}[t!h!]
	\epsfxsize=\columnwidth
	\begin{center}
		\epsfbox{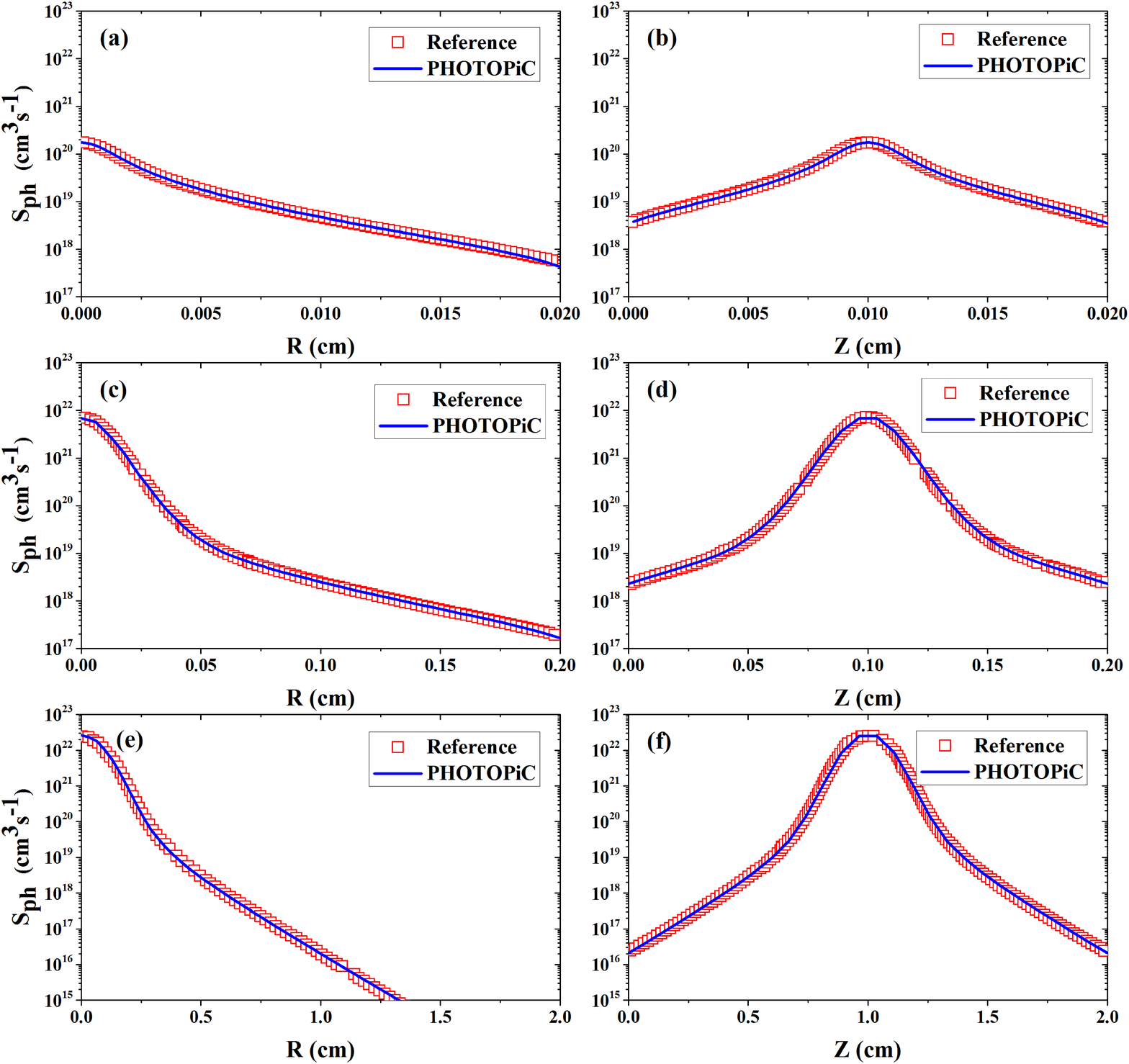}	
	\caption{Photo--ionization source terms calculated by PHOTOPiC and in Ref~\cite{Bourdon2007}. (a) and (b) correspond to the radial and axial distribution of photo--ionization source terms in 0.02~cm$\times$0.02~cm scale. (c) and (d) correspond to 0.2~cm$\times$0.2~cm scale while (e) and (f) correspond to 2~cm$\times$2~cm. The red lines are calculated by extended three terms Helmholtz model using the parameters given by PHOTOPiC, the dots are extracted from the reference.}
\label{validation2}
	\end{center}
\end{figure}

The calculated photo--ionization source term required for plasma modeling is shown in Figure~\ref{fig.source}. The distribution of photo--ionization source terms in radial and axial directions calculated from the classical three terms Helmholtz model are plotted together for comparison. It is clearly seen that, the extended Helmholtz photo--ionization model is equivalent to the classical one that is designated for $\rm N_2/O_2$ mixtures.

\subsection{Streamer propagation in $\rm N_2/O_2$ mixture}

The parameters calcualted for the extended three terms Helmholtz model are used in a 2D fluid code (PASSKEy code\cite{zhu2017nanosecond}) to reproduce the results of a classical streamer discharge simulation benchmark~\cite{Kulikovsky1998}. This benchmark was done for the hyperboloid anode placed 1 cm over a plane cathode in atmospheric pressure air. Constant voltage of 13 kV was applied to the anode. An accurate numerical scheme to correctly capture the avalanche--to--streamer transition and the propagation velocity. 

\begin{figure}[t!h!]
	\epsfxsize=\columnwidth
	\begin{center}
		\epsfbox{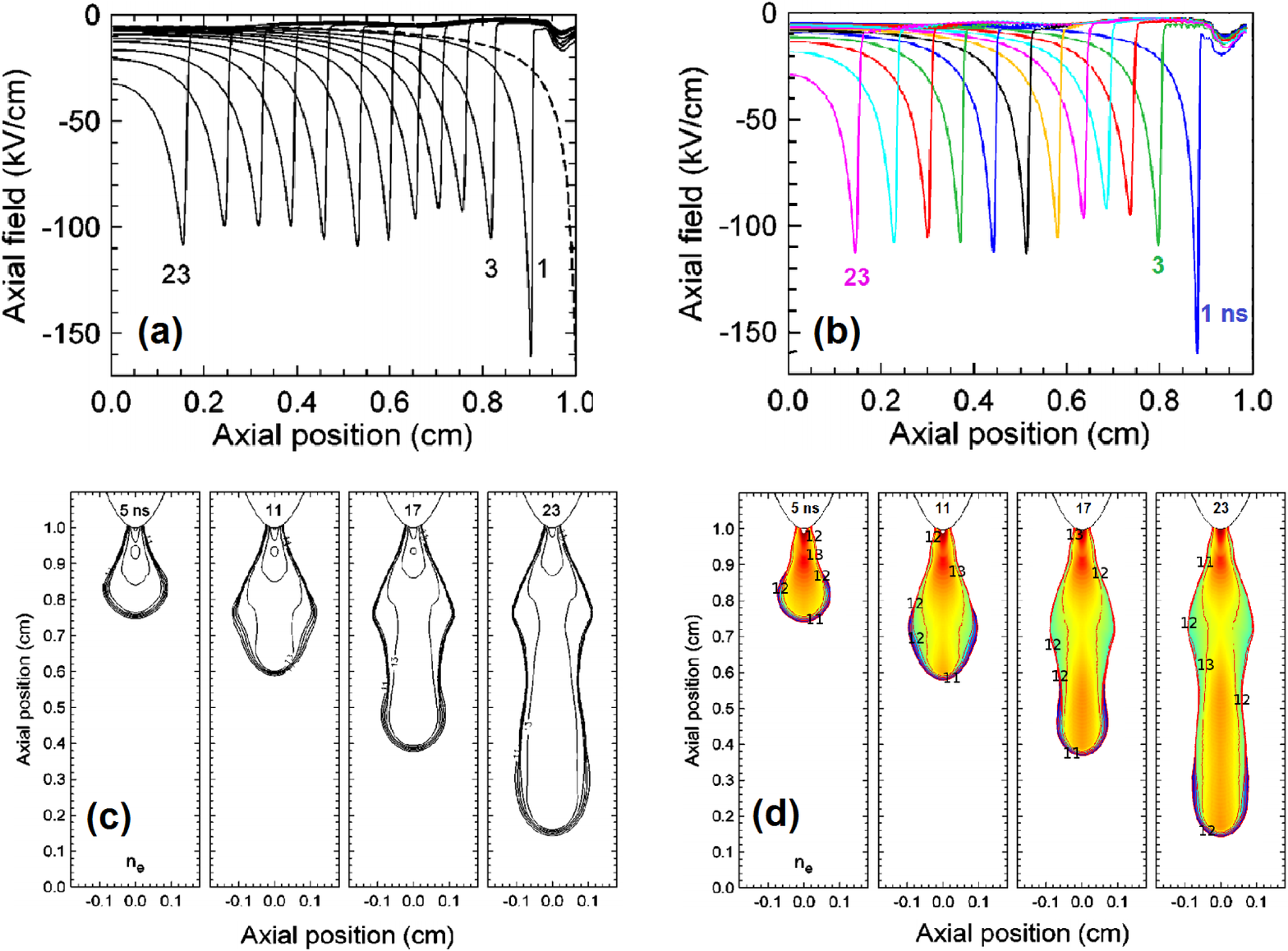}	
	\caption{The pin--plane discharge benchmark. (a) and (c): the evolution of axial electric field and spatial electron density in Ref~\cite{Kulikovsky1998}; (b) and (d) results calculated by PASSKEy code~\cite{zhu2017nanosecond} using the parameters obtained by PHOTOPiC.}
\label{validation3}
	\end{center}
\end{figure}

Figure~\ref{validation3}~(a-d) compares the results of the calculation of the present work with the reference results taken from \cite{Kulikovsky1998}. Calculated in the present work axial profiles of the electric field and isolines of the electron density are shown in Figure~\ref{validation3}~b and d respectively. 

Different computational approaches to photoionization have been used in \cite{Kulikovsky1998} and in the present work. Both codes are based on the classical photoionization model developed by Zheleznyak et al~\cite{Zhelezniak1982}. In \cite{Kulikovsky1998}, the photoionization was calculated by integration over the region containing the emission sources; the region was a restricted volume related to the streamer head. The present work uses the extended three terms Helmholtz model. Despite the described difference, the results calculated by the PASSKEy code provide a good agreement with \cite{Kulikovsky1998}, see Figure~\ref{validation3}~a and c. The streamer is initiated in high Laplacian field close to the anode and expands along the axis of
the discharge and in the radial direction until it reaches approximately a radius of
1 mm. After this, the streamer propagates along the axis of the discharge gap with
almost constant radius of the channel and practically constant velocity.

\section{Conclusions}

A program devoted for calculating the photo--ionization functions is developed. The calculated photo--ionization functions have been validated by existing experimental data and can be used in different numerical models for gas discharges. 
The classical three terms Helmholtz photo--ionization model is extended to account for photo--ionizations without oxygen. We have validated that the extended three terms Helmholtz photo--ionization model is equivalent to the classical one for $\rm N_2/O_2$ mixtures.

Future versions of the program will provide the possibility to add user--defined gas specie, and the capability to calculate parameters for other photo--ionization models (i.e. the three-group Eddington and SP$_3$ approximation model et al.). 

\section*{Acknowledgements}
The work was supported by the National Natural Science Foundation of China (No. 51907204, 51790511, 91941105, 91941301) and the National Numerical Windtunnel Project NNW2018-ZT3B08. The authors are thankful to the young research group in Atelier des Plasmas for fruitful discussions.





\bibliographystyle{elsarticle-num}
\bibliography{PHOTOPic_Lib}







\end{document}